\documentclass[a4paper,11pt]{article}

\usepackage{jinstpub} 
\usepackage{threeparttable}

\title{LTARS: Analog Readout Front-end ASIC for Versatile TPC-applications}

\author[a,1,2]{T. Kishishita,\note{Corresponding author.}\note{Also at SOKENDAI, The Graduate University for Advanced Studies.}}
\author[b]{S. Sumomozawa,}
\author[b]{T. Kosaka,}
\author[b]{T. Igarashi,}
\author[a]{K. Sakashita,}
\author[a]{M. Shoji,}
\author[a]{M. M. Tanaka,}
\author[a]{T. Hasegawa,}
\author[b]{K. Negishi,}
\author[b]{S. Narita,}
\author[c]{T. Nakamura}
\author[c]{, and K. Miuchi}

\affiliation[a]{KEK, High Energy Accelerator Research Organization\\1-1 Oho, Tsukuba, 3050801, Ibaraki, Japan}
\affiliation[b]{Iwate University,\\4-3-5 Ueda, Morioka, 0208551, Iwate, Japan}
\affiliation[c]{Kobe University,\\1-1 Nada-ku, Rokkodaicho, Kobe, 6570013, Hyogo, Japan}
\emailAdd{kisisita@post.kek.jp}

\newcommand{\ie}{\textit{i.e.}}
\newcommand{\eg}{\textit{e.g.}}

\abstract{
We designed a versatile analog front-end chip, called LTARS, for TPC-applications, primarily targeted at dual-phase liquid Ar-TPCs for neutrino experiments and negative-ion $\mu$-TPCs for directional dark matter searches. Low-noise performance and wide dynamic range are two requirements for reading out the signals induced on the TPC readout channels. One of the development objectives is to establish the analog processing circuits under low temperature operation, which are designed on function block basis as reusable IPs (Intellectual Properties). The newly developed ASIC was implemented in the Silterra 180~nm CMOS technology and has 16 readout channels. We carried out the performance test at room temperature and the results showed an equivalent noise charge of 2695$\pm$71~e$^-$ (rms) with a detector capacitance of 300~pF. The dynamic range was measured to be 20--100~fC in the low-gain mode and 200--1600~fC in the high-gain mode within 10\% integral nonlinearity at room temperature. We also tested the performance at the liquid-Ar temperature and found a deterioration of the noise level with a longer shaper time. Based on these results, we also discuss a unique simulation methodology for future cold-electronics development. This method can be applicable to design the electronics used at low temperature. 
}

\keywords{Time projection chambers, Front-end electronics for detector readout, Liquid detectors}

\begin{document}
\maketitle
\flushbottom

\section{Introduction}
In recent studies in particle and astro-particle physics, detectors with high-resolution positional imaging are often desired to explore new physics phenomena because the topological information can be utilized to discriminate the signature of the new physics phenomena from background events. These detectors require huge numbers of readout channels and advanced electronics play an important role in handling such large numbers of readout channels. 
To meet this need, we are developing front-end electronics for versatile applications of time projection chambers (TPCs) for a joint project  called {\it LTARS} (Low Temperature Analog Readout System), located at KEK, Kobe, and Iwate
Universities. We plan to use this readout system for 
the directional dark matter experiments \cite{NEWAGE,CYGNUS} 
and for the next-generation neutrino oscillation experiments \cite{nuexp}. 

\begin{figure}[htbp]
 \begin{minipage}{0.5\hsize}
  \begin{center}
   \includegraphics[width=75mm]{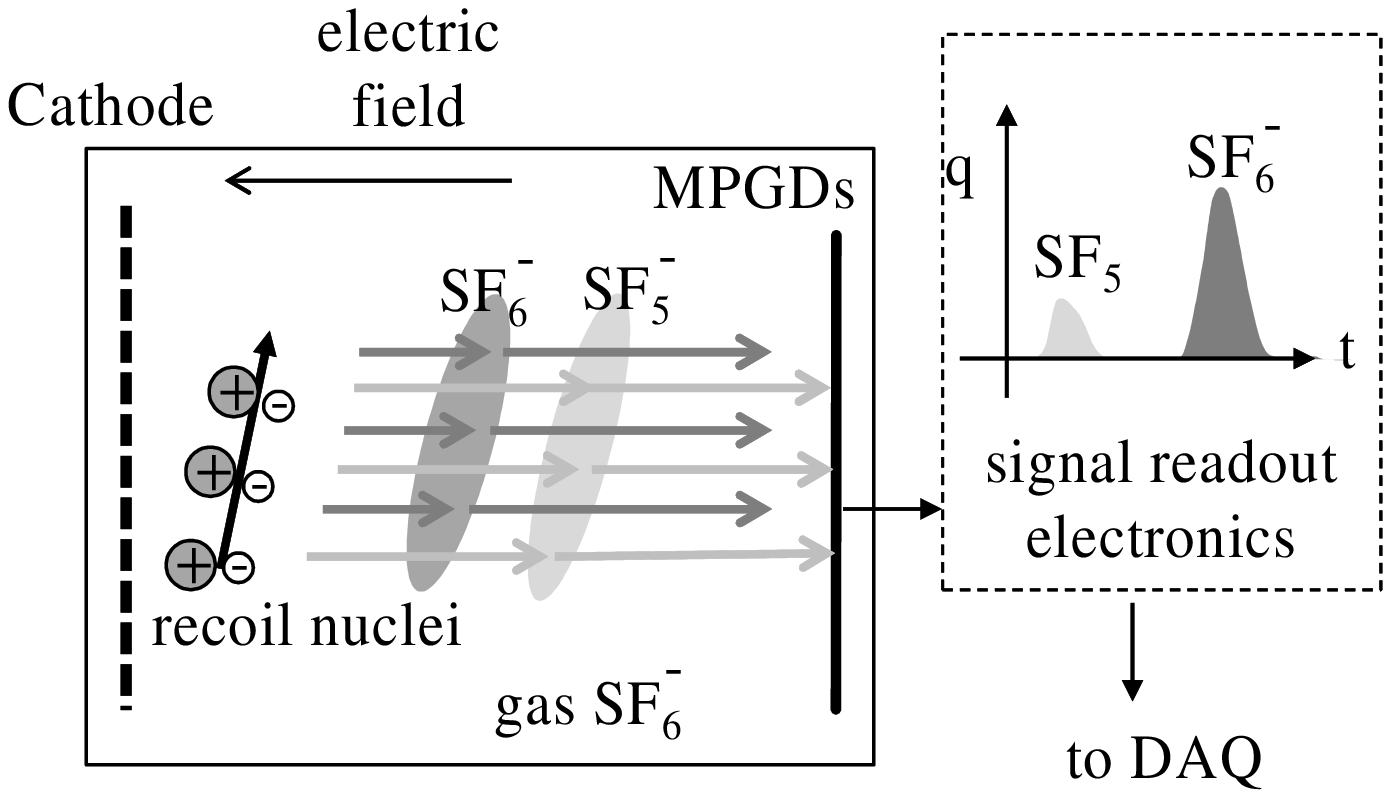}
  \end{center}
 \end{minipage}
 \hspace{0.5cm}
 \begin{minipage}{0.45\hsize}
  \begin{center}
   \includegraphics[width=60mm]{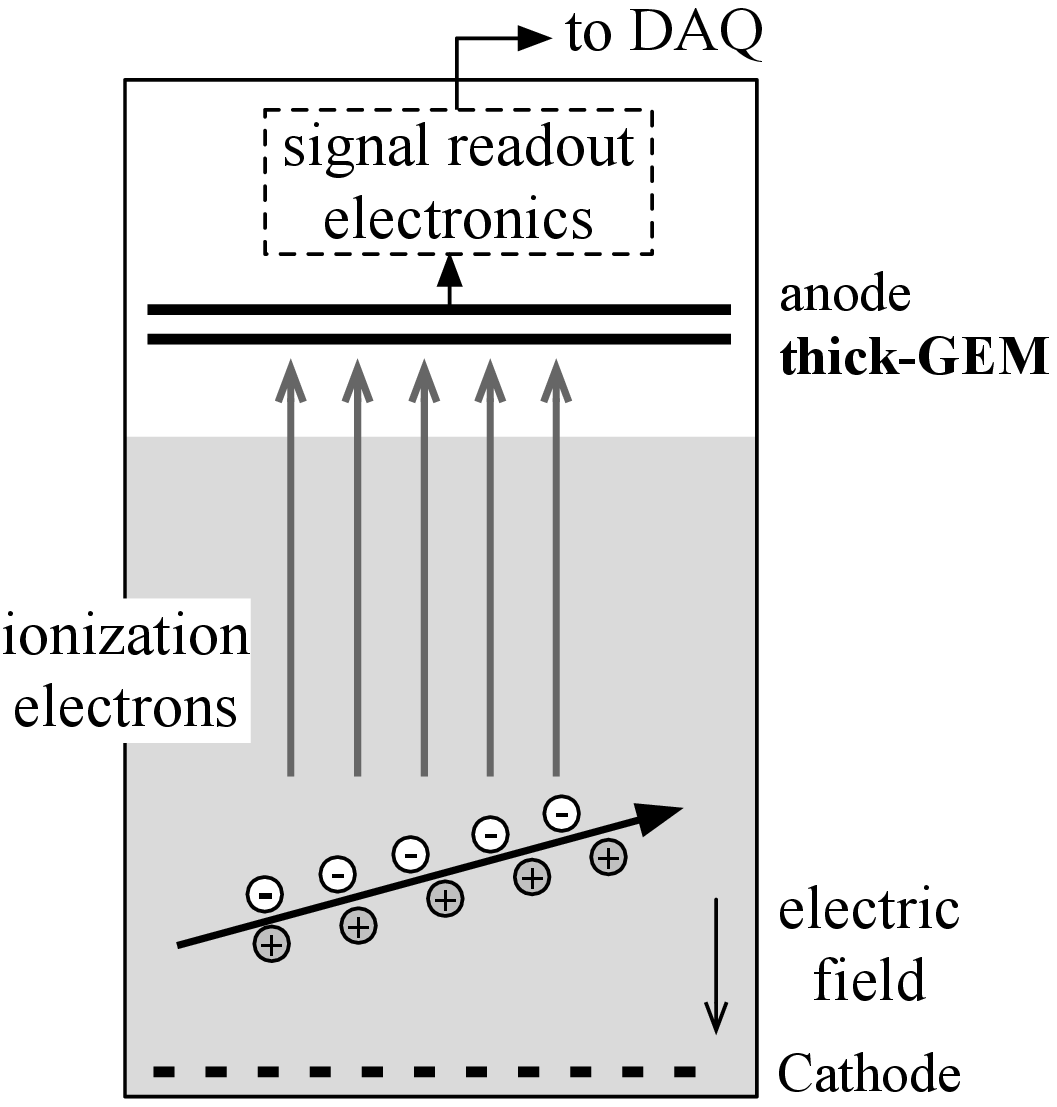}
  \end{center}
 \end{minipage}
  \caption{Detector concepts of the NI$\mu$-TPC (Left) and the dual-phase LAr-TPCs (Right). }
  \label{fig:TPC}
\end{figure}

TPC-based three-dimensional (3D) tracking detectors are thought to be one of the best detectors for the directional direct dark matter search experiments and several groups have developed prototype detectors \cite{tanimori, mayet}. 
{\it NEWAGE} is one of these directional dark matter search experiments and has been leading in directional sensitivity \cite{NEWAGE}. 
One of the next steps needed to improve the sensitivity is to utilize a negative-ion drift gas, which enables 
full-volume fiducialization through 3D position reconstruction of the event vertex \cite{minority,SF6_Ikeda2020}. 
These TPCs are called negative-ion (NI) $\mu$-TPCs and the detector concept is shown in the left of Figure \ref{fig:TPC}.
In the NI $\mu$-TPCs, the ionized electrons are captured by the gas molecule immediately after ionization thereby generating negative ions. These ions are drifted instead of the electrons in the NI $\mu$-TPCs.  Some types of negative ion gases could contain 
 more than one species of negative ion ($\rm SF_6$ and $\rm SF_5$ in Figure \ref{fig:TPC}). Each of these ions are drifted with different velocities. The arrival time difference provides information on drift length, or, the absolute position on the electric field direction. 
This new concept helps to improve the sensitivity by rejecting background events from the readout plane (indicated as MPGDs in Figure \ref{fig:TPC}) and the cathode plane; this was not possible with the self-triggering TPCs.

A large scale $\mathcal{O}$(10~kt) liquid argon time projection chamber (LAr-TPC) will be utilized as a 3D-tracking device for studies of next-generation neutrino oscillation, nucleon decay, and astrophysical neutrinos. 
Toward the realization of such a large-scale detector,
a world-wide R\&D effort on kilo-ton-scale LAr-TPC demonstrators is underway \cite{WA105}.
One approach to read out the ionized electron signals is to use a dual-phase LAr-TPC \cite{LEM}.
The detector concept is shown in the right of Figure~\ref{fig:TPC}.
In the dual-phase TPC, the ionized electrons are extracted from the liquid phase (indicated as LAr in  Figure~\ref{fig:TPC}) to the gas phase (GAr), 
and the extracted electrons are multiplied with a thick-GEM (Gas Electron Multiplier) and collected with a two-dimensional strip anode. 
The main advantage of the dual-phase readout is the high signal-to-noise ratio afforded 
by the gas multiplication. This enables a longer drift length because the signal is amplified even though some primary electrons are lost by impurities in the LAr.
The high signal-to-noise ratio also benefits the 
physics sensitivity for the neutrino oscillation, nucleon decay, and astrophysical neutrino signals.

The signal timescale is much longer than that of electrons in typical gas TPCs, \ie{},$\approx 10^{-2}$ cm/$\mu$s for the NI $\mu$-TPCs and $\approx 10^{-1}$ cm/$\mu$s for the LArTPCs, respectively. Therefore, the electronics has a lot of similarities, but is different from the standard electronics for the TPC.
Additionally, a wide dynamic range together with a high signal-to-noise ratio is required for the readout electronics. 
In the case of the NI $\mu$-TPCs, 
typical signals are 3~fC for the minority species (indicated as $\rm SF_5$ in  Figure~\ref{fig:TPC}) and 80 fC for the main species ($\rm SF_6$) \cite{miuchi, giomataris}. 
On the other hand, the typical ionization signal in the LArTPCs is $\approx 10$~fC, assuming one order of magnitude of the thick-GEM gains, for a minimum ionization particle (MIP). 
The signal on one readout channel 
for events associated with electromagnetic or hadronic showers is about 50 times larger than 1 MIP signals. 
The detector capacitances seen from the ASIC input are both estimated as $\sim 300$~pF. 
A fine spatial resolution is also necessary for both detectors, and it results in a huge number of readout channels as a whole detector system. We also need to take the operating temperatures into consideration. 

Whereas the readout electronics can be operated at room temperature (RT) in the NI$\mu$-TPC, the front-end needs to be operated at $\approx$-185~$^\circ$C for in the LArTPCs. This thermal constraint comes from a practical issue that the analog front-end electronics must be located as close as possible to the strip readout in order to minimize the detector capacitance.

To satisfy various requirements from both experiments, we have developed a series of readout ASICs in a 180 nm CMOS technology. The circuits are designed on a function-block basis as reusable components. In this paper, we report on the performance of a newly developed 16-channel ASIC at RT and the functionality under the LAr temperature (LT) operation.
We optimized the transistor parameters of the previous prototype chip, taking special care in the layout to improve low-noise characteristics \cite{nakazawa}. One of the objectives of LTARS development was to generate  cold-electronics 'know-how' and to clarify issues for future system-integration, \eg{}, operating methods of the ASICs at LT, and geometrical constrains on readout systems for a huge number of channels. We describe the ASIC design in Section 2, report on its performance at RT and LT in Sections 3 and 4, respectively, discuss our simulation methodology for cold-electronics and future developments in Section 5, and conclude in Section 6.
 

\section{Architecture of the ASIC}
\subsection{Circuit properties}
\begin{figure}[htbp]
  \begin{center}
   \includegraphics[width=40mm]{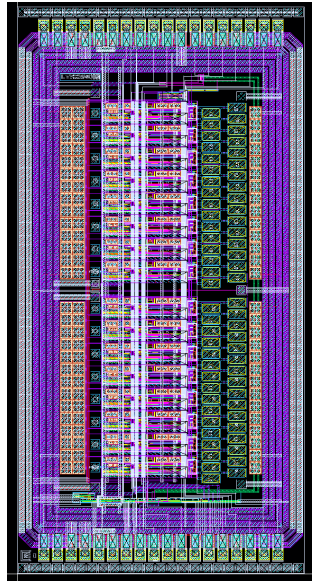}
  \end{center}
  \caption{Physical layout of the readout ASIC. The chip size is 2.5~mm$\times$5~mm.}
  \label{fig:two}
 \end{figure}

\begin{figure}[htbp]
\begin{center}
\includegraphics[width=140mm]{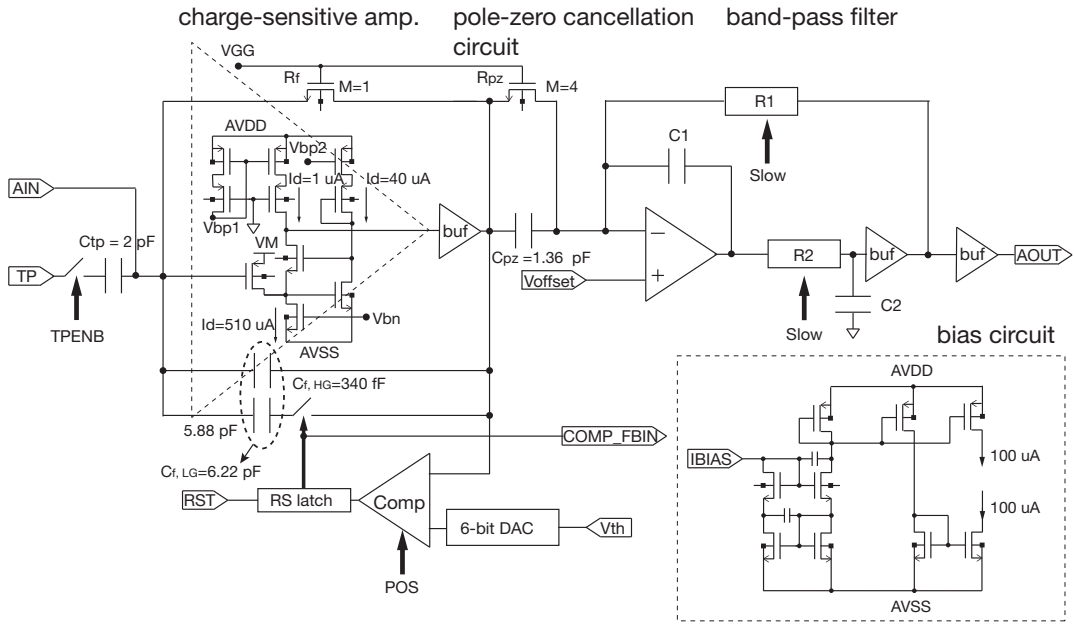}
\end{center}
\caption{Block diagram of the processing chain. Unlabeled substrates are connected to the supply powers.}
\label{fig:circuit}
\end{figure}

The physical layout of the readout ASIC implemented in the Silterra 180~nm CMOS technology is shown in Figure \ref{fig:two}.  Technological parameters and requirements for the ASIC are listed in Table \ref{tab1}. The chip includes 16 identical signal processing channels. The block diagram of each channel is shown in Figure \ref{fig:circuit}. The TPC readout channel is connected to an input (AIN) of the charge-sensitive amplifier (CSA) by an off-chip capacitor, while test pulses can be injected via an on-chip AC-coupling capacitor $C_{\rm tp}=2$~pF. The CSA is based on a folded-cascode  configuration with a p-channel input transistor and a regulated-cascode configuration is implemented to improve the open-loop gain for large detector capacitances. A transfer-gate type FET is employed for the CSA DC-feedback component. Two distinct feedback capacitors are implemented for dynamic gain-switching by using a metal-insulator-metal structure. The feedback capacitor is initially set at $C_{\rm f,~HG}=340$~fF. This state corresponds to a high-gain (HG) mode and it copes with a narrow range signal of $<80$~fC. The CSA output is fed into a discriminator to select the gain mode.
Once the amplitude of the CSA exceeds a certain threshold, which is given by a 6-bit DAC, a discriminator followed by an RS-type flip-flop latches the switch on the additional feedback capacitance. This state functions as a low-gain mode (LG) with a capacitance value of $C_{\rm f,~LG}=6.22$~pF. As a result, the overall voltage gain is more than 10 times smaller than that in the HG mode. After reading out the analog output (AOUT) and the gain information (COMP\_FBIN), the reset signal (RST) supplied externally releases the latched signal and the overall circuit returns to the idle state.
The influence of the switching noise on the noise performance is negligible compared to the large input signals required to switch on the LG mode. The dynamic-switching behaviors have been demonstrated in the prototype chip, and this unique property makes the chip multi-purpose, not only for LAr-TPCs but also NI $\mu$TPCs \cite{nakazawa}.

A CR-RC band-pass filter is composed of a pole-zero cancellation circuit (PZC) and a second-order integration low-pass filter. The capacitance and resistance values were selected to meet the equation of $\displaystyle{ C_{\rm f,~HG} \cdot R_{\rm f}=C_{\rm pz}\cdot R_{\rm pz}}$, and $C_1\cdot R_1=4\times C_2\cdot R_2$ (see Fig. \ref{fig:circuit}).
The transfer functions in the HG mode are described as \cite{geronimo}
\begin{eqnarray}
T_{\rm CSA}&=&-Q_{\rm in}\cdot \frac{R_{\rm f}}{(1+sC_{\rm f,~HG}\cdot R_{\rm f})}\cdot \frac{(1+sC_{\rm pz}\cdot R_{\rm pz})}{R_{\rm pz}},\nonumber  \\
T_{\rm CR-RC}&=& -\frac{R_1}{(2sC_2\cdot R_2+1)^2}, \nonumber \\
T_{\rm total}&=&Q_{\rm in}\cdot \frac{R_{\rm f}\cdot R_1}{R_{\rm pz}}\cdot \frac{1}{(2sC_2\cdot R_2+1)^2},
\end{eqnarray}
where $s$ denotes the complex angular frequency and $Q_{\rm in}$ is the input charges. 
The default shaping-time of the CR-RC filter is designed as 1~$\mu$s for the LArTPCs. This value can be switched to 4~$\mu$s for NI $\mu$TPCs via a 9-bit control register. This register is equipped in each channel, while also controlling the test pulse enable, monitor enable, and tuning voltage threshold. 
A reference current is injected to the IBIAS node via a potentiometer inserted between the ground and supply power. The current-mirror configuration generates an internal bias current of 100~$\mu$A that provides the proper bias current to each circuit block, \eg, 510~$\mu$A for the input FET in the CSA, at the current step of 10~$\mu$A.

 \begin{table*}[!htbp]
\begin{threeparttable}
\caption{\label{tab1} Technological parameters and requirements to the ASIC.}
\def\arraystretch{1}
\ignorespaces 
\centering 
\small
  \begin{tabular}{|c|c|c|} \hline
Technology & \multicolumn{2}{|c|}{Silterra 180~nm CMOS} \\ \hline 
Chip size & \multicolumn{2}{|c|}{2.5$\times$5~mm$^2$} \\ \hline
The number of channels &  \multicolumn{2}{|c|}{16} \\ \hline
Supply power & \multicolumn{2}{|c|}{1.8~V core/IO, max. 2.4~mW/ch}\\ \hline
Fabrication options & \multicolumn{2}{|c|}{6 metals, deep N-well, high-value poly res., MIM cap.} \\ \hline \hline
Detector type& \hspace{10ex} NI$\mu$-TPC  \hspace{10ex}  & LAr-TPC \\ \hline
Minimum signal charge & $\approx$3 fC & $\approx$10 fC  \\ \hline
Shaping time & 4~$\mu$s & 1~$\mu$s \\ \hline
Operating condition & room temperature & -185~$^\circ$C \\ \hline
Detector capacitance (C$_{\rm det}$) \tnote{a}  & \multicolumn{2}{|c|}{$\sim$300~pF} \\ \hline
Dynamic range & \multicolumn{2}{|c|}{$\pm$80 fC  for narrow range, $\pm$1600 fC for wide range}\\ \hline
Voltage gain & \multicolumn{2}{|c|}{10~mV/fC for narrow range, 0.5~mV/fC for wide range}\\  \hline
ENC & \multicolumn{2}{|c|}{3000~e$^-$ (S/N$>$20) for small signals, $< 6.4 \times 10^4$ e$^-$ for large signals}\\ \hline
\end{tabular}
\begin{tablenotes}
\item[a] Estimated from the pad size of MPGDs.
\end{tablenotes}
\end{threeparttable}
\end{table*}

\section{Performance test results at room temperature}

 \subsection{Experimental setup}
\begin{figure*}[htbp]
\begin{center}
\includegraphics[width=150mm]{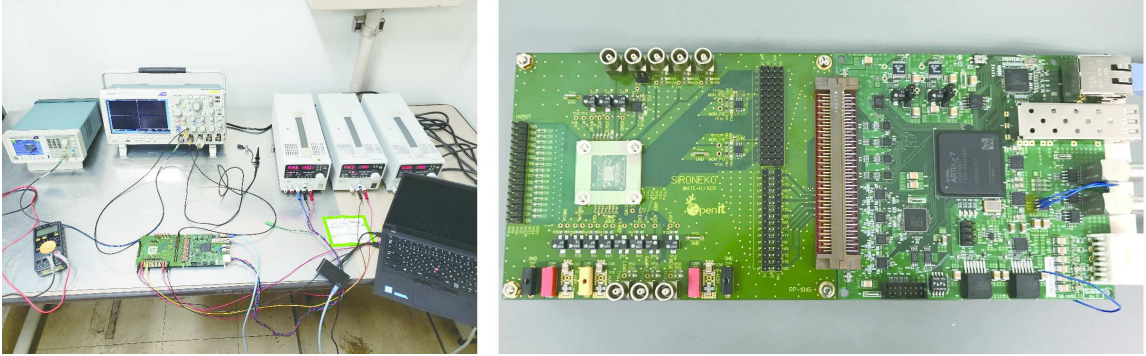}
\end{center}
\caption{(Left) Experimental setup at room temperature. (Right) Testboards of the ASIC.}
\label{fig:four}
\end{figure*}

The experimental setup and the dedicated  printed circuit board (PCB) for performance testing are show in Figure \ref{fig:four} left and right, respectively. The chip-mounted board (named SIRONEKO) shown on the left side of the right figure provides electrical connections between the ASIC. The semi-custom FPGA board (named GoSHIK) is also shown on the right side. A bare die was directly mounted on the PCB, and optical light was shielded during the measurements. The GoSHIK board is an interface with a computer and provides the register control signals.
This board includes 8-channel ADCs (AD9637) and 8-channel voltage/current DACs (LTC2656 and MAX5550). It provides a flexible bias setting to the ASIC, along with an easy-to-use pattern generation from the FPGA. We chose the Xilinx XC7A100T-2FGG676C \cite{FPGA}, and data transfer is done via an Ethernet cable with the SiTCP protocol \cite{SITCP}. Test pulses are generated by a function generator (AFG-21025).

\subsection{Analog waveforms and dynamic range}
\begin{figure}[htbp]
  \begin{center}
   \includegraphics[width=150mm]{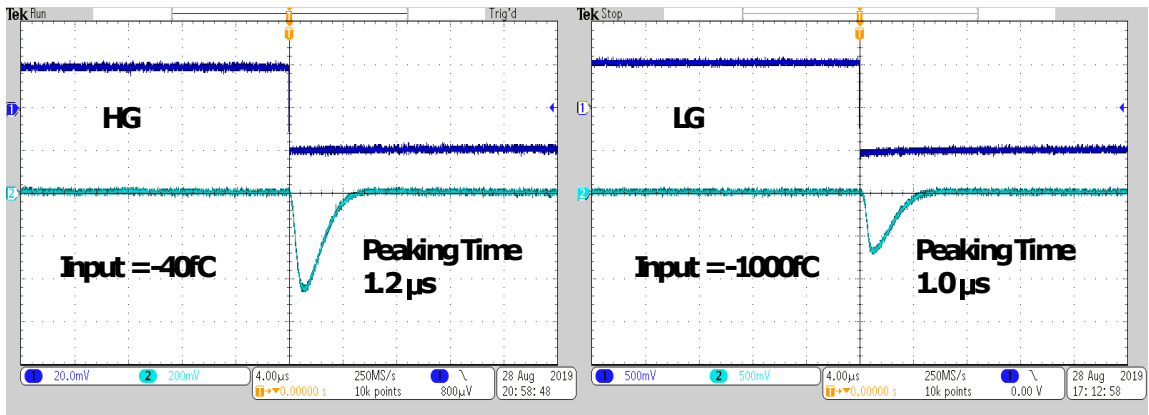}
  \end{center}
  \caption{Analog outputs. The intervals on the vertical axis are 20~mV for input and 200~mV for output in the HG mode, and 500~mV for input and output in the LG mode. The intervals on the horizontal axis are 4 $\mu$s. All measurements were performed under the condition of $C_{\rm det}$=0~pF.}
  \label{fig:five}
 \end{figure}

We first injected test pulses and checked the analog output with an oscilloscope. Figure \ref{fig:five} shows the waveforms in the different gain modes. The gain-switch and shaping time (1~$\mu$s) were fixed during the measurements.
 The waveforms were obtained without detector capacitance. Test pulses are shown in blue, which correspond to input charges of -40 fC in the HG mode and -1000 fC in the LG mode. The peaking times were measured as  1.2~$\mu$s and 1.0~$\mu$s, respectively. Although the pole and zero in the transfer function $T_{\rm CSA}$ are not cancelled out in the LG mode (see Eq. (2.1)), the overshoot was negligible at the analog output. The difference of the peaking times is not a major issue as long as the analog outputs are continuously sampled in parallel by ADCs, and the waveform is reconstructed in offline analysis.
 
\begin{figure}[htbp]
  \begin{center}
   \includegraphics[width=140mm]{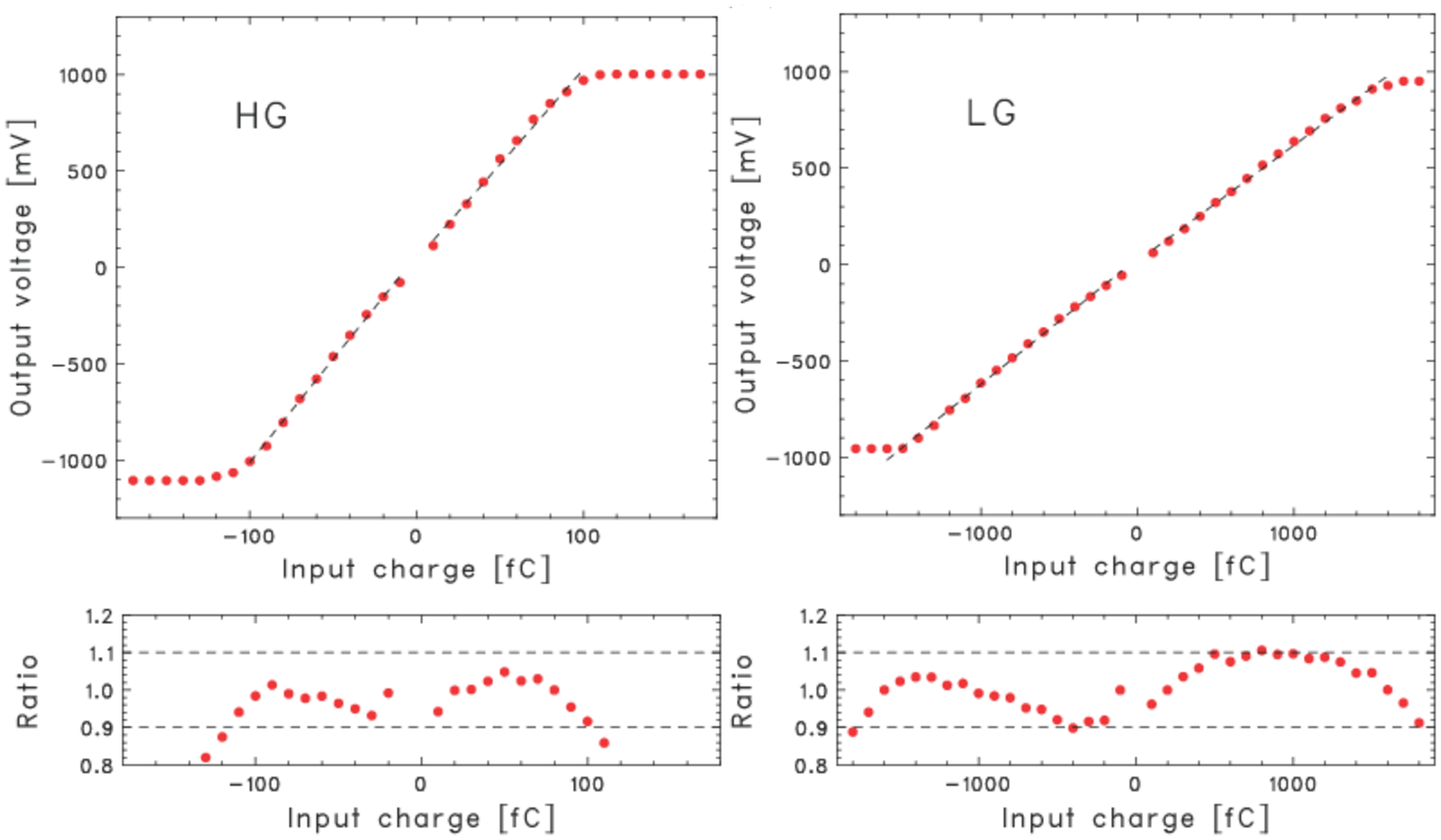}
  \end{center}
  \caption{Dynamic ranges of a typical readout channel in different gain modes. The lower panels show the residual between the data and linear functions. The baseline is subtracted from the peak pulse height. }
  \label{fig:six}
 \end{figure}

\begin{figure}[htbp]
  \begin{center}
   \includegraphics[width=100mm]{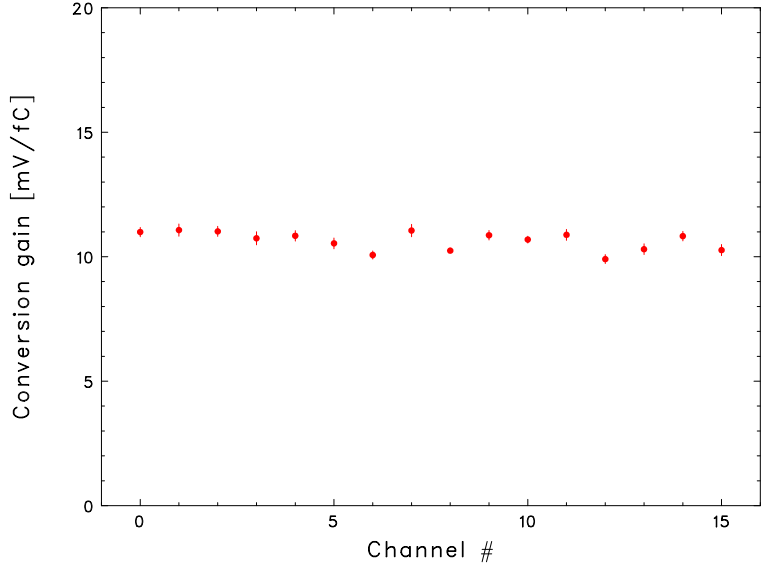}
  \end{center}
  \caption{Gain variation in the HG mode. }
  \label{fig:gainvsch}
 \end{figure}

The dynamic range of a typical channel is shown in Figure \ref{fig:six}. 
Conversion gains were obtained by fitting a line through the minimum and maximum points of the required dynamic range. The measured values were 10.0 mV/fC for positive and 10.7 mV/fC for negative polarity in the HG mode, whereas in the LG mode, these values were 0.60 mV/fC and 0.65 mV/fC, respectively . 
The lower panels show the residuals between the data and fitting lines. Linearity is maintained with $\pm$ 10\% integral nonlinearity up to $\pm$100 fC in the HG mode, while that extends to $\pm$1600 fC in the LG mode. 
The gain variation between the 16 channels is shown in Figure \ref{fig:gainvsch}. We confirmed that the 
variation is within 10\%.
 
 \subsection{Equivalent noise charge}
\begin{figure}[htbp]
  \begin{center}
   \includegraphics[width=100mm]{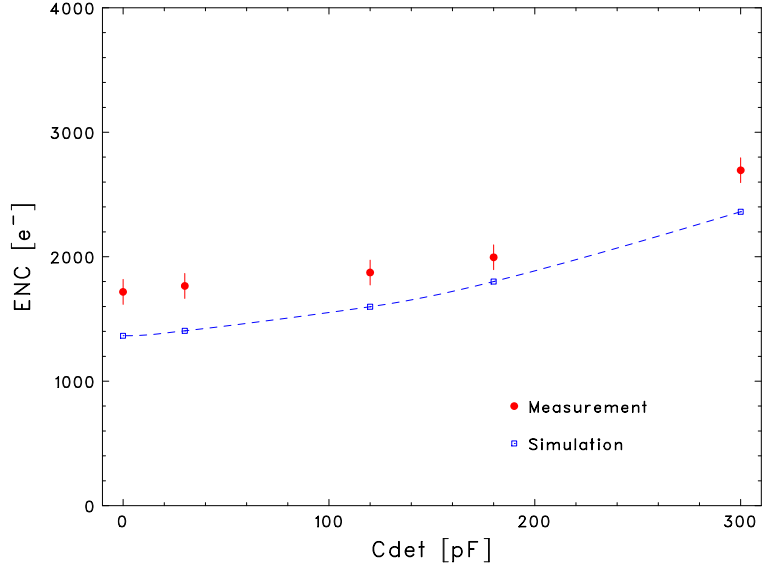}
  \end{center}
  \caption{The equivalent noise charge (rms) as a function of detector capacitance. Data and simulation are shown in red and blue, respectively.}
  \label{fig:seven}
 \end{figure}

All the noise generated inside the amplifier is calculated as a quantity that is then converted into an equivalent noise charge (ENC), which is the noise generated at the input.
The ENC is given by the following formula:
 \begin{equation}
{\rm ENC~(electron) = \frac{V_{\rm noise, rms}~(mV)}{Conversion\ gain (mV/fC) \times 1.6 \times 10^{-4}(fC)}}. \nonumber \\
\end{equation}
Figure \ref{fig:seven} shows the detector capacitance $C_{\rm det}$ versus the corresponding ENC values. For comparison, the simulation result is overlaid in the figure. To improve the noise performance, on-chip ESD protection diodes were not included in the analog inputs. 
The voltage noise is proportional to the detector capacitance, and the ENC in the HG mode was measured as 2695$\pm$71~ $e^{-}$ at $C_{\rm det}=300$~pF. By comparing our results with the simulation value of  2361 $e^{-}$, we can confirm that the performance is very close to the simulation value, although there is about 13\% offset at $C_{\rm det}=300$~pF. Possible causes of the noise offset are the bonding wire and PCB trace capacitance due to the mounting of the ASIC on the evaluation board and the ground bounce due to a single-supply configuration, \ie, 0/1.8~V. Based on these considerations, we expect to be able to improve the performance of the ASIC in the next experiment and design.  
The ENC in the LG mode was measured as 37200~$\pm$330 $e^{-}$ at $C_{\rm det}=300$~pF with a noise slope of 5.41 ~e$^-$/pF. Since the expected value is 36913~$e^{-}$, we concluded that the overall performance at RT are in agreement with the simulation models provided by the vendor. Moreover, the measured value is lower than the requirement of 64000 $e^{-}$ in the LG mode.

The typical ionization signal for 1 MIP in the LAr-TPC of 10 fC, corresponding to 94000$e^{-}$, is expected based on recent studies at the large LAr-TPC demonstrators \cite{WA105}. The ENC of 2700~$e^{-}$achieves S/N = 23, and thus, the ASIC performance is considered to have reached the level of practical application if the RT performance can be kept at the LAr temperature. The data processing architecture in LArTPCs to handle the analog outputs from all channels is currently under discussion. Two options exist: either using external ADCs in parallel or on-chip ADCs combined with sparse readout. Since the number of readout channels is expected to be $\mathcal{O}(10^5)$ for a $\mathcal{O}(10\,{\rm kt})$ detector, it is desirable to reduce the number of feedthrough lines running from the inside to the outside of the LAr cryostat.  From this perspective, we will consider a circuit that consists of both analog and digital processing parts by optimizing the deep Nwell option of the current 180$\,$nm CMOS technology.

\section{Measurements at the LAr temperature}
\subsection{Experimental setup}

\begin{figure}[htbp]
 \begin{minipage}{0.5\hsize}
  \begin{center}
   \includegraphics[width=70mm]{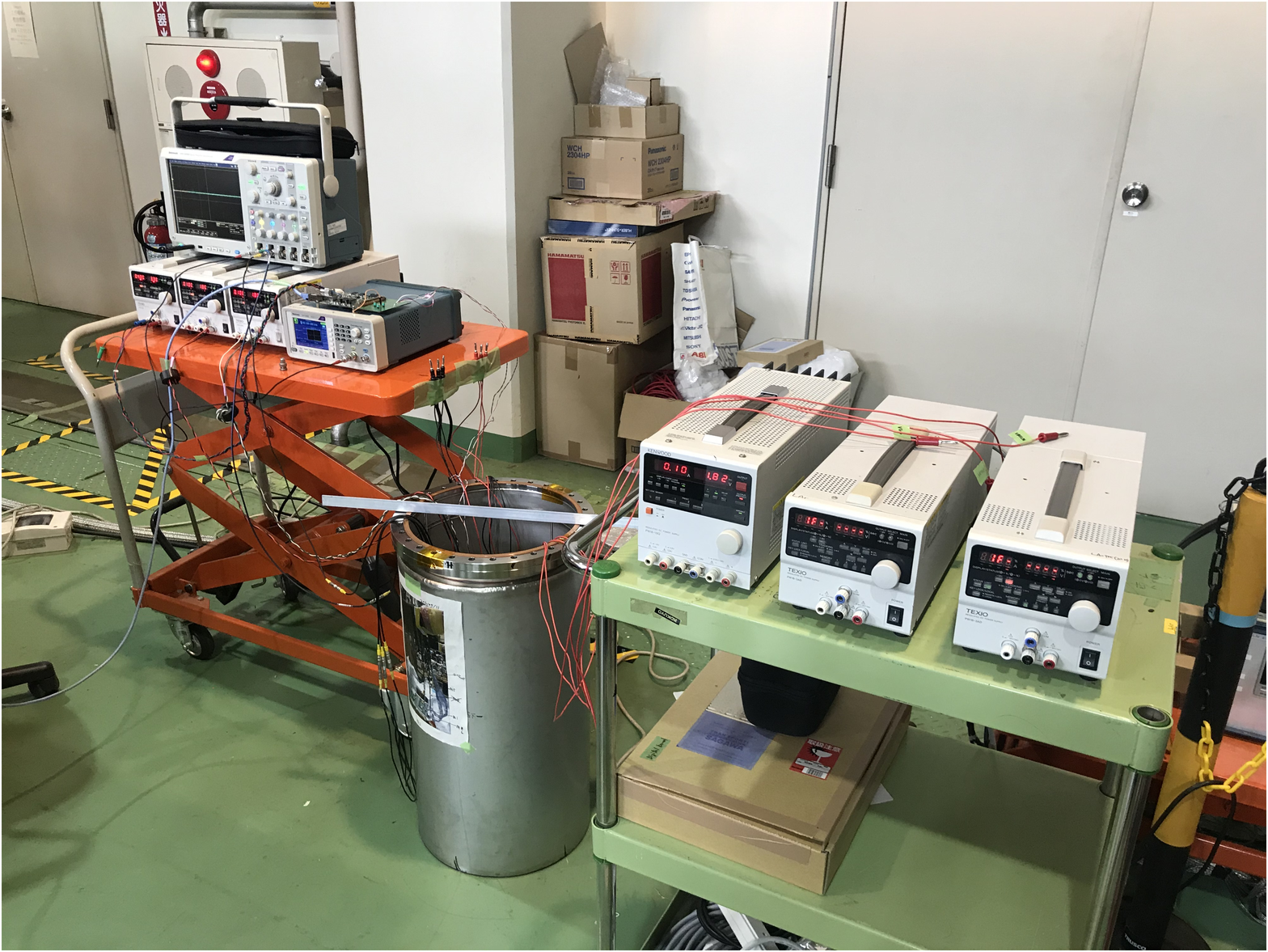}
  \end{center}
 \end{minipage}
 \begin{minipage}{0.5\hsize}
  \begin{center}
   \includegraphics[width=70mm]{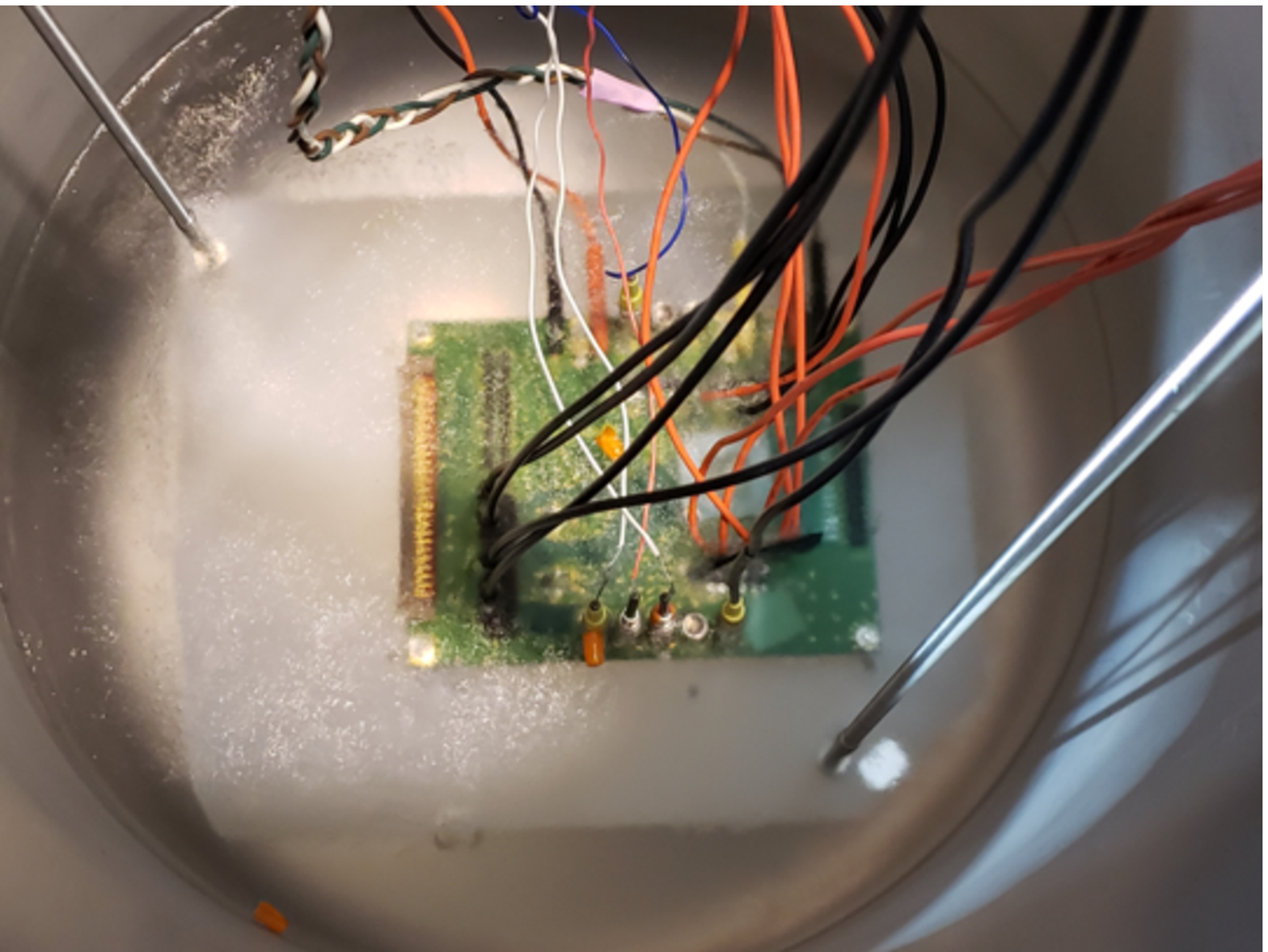}
  \end{center}
 \end{minipage}
  \caption{Experimental setup for the LAr temperature operation (left). Only the ASIC, \ie, the SIRONEKO board, was directly immersed in the LAr (right).}
  \label{fig:eight}
\end{figure}

Figure \ref{fig:eight} shows the experimental setup operated at the LAr  temperature of $\approx$-185~$^\circ$C. Since the feed-through terminal of the cryostat limits the number of cable connections, we chose direct immersion of the electronics in a Dewar vessel filled with liquid argon. To avoid thermal stress on the FPGA board, we separated the GoSHIK board from the ASIC, soaking only the SIRONEKO board in liquid argon. Supply powers, test pulses, and monitor lines were directly connected with ribbon cables. The bias voltages, which were tuned with trimmer potentiometers at RT, were also provided by external power supplies. The direct immersion approach provides an easy-to-access environment to the ASIC, although is subjects the ASIC and mounted components to harsh thermal and mechanical stresses. In this experiment, the temperature can be reduced to that of liquid argon instantaneously, removing any  time restrictions.

\subsection{Waveforms and dynamic range}
In the LAr temperature, we confirmed that the circuit could not be operated under the same bias conditions as the RT environment. By optimizing the bias settings we succeeded in obtaining analog outputs; however, the conversion gain decreased and the noise level severely deteriorated. Such deterioration was not observed at RT in the Dewar. 
The baseline fluctuation was about 120~mV, while the peak height was about 530mV for an input charge of 40~fC. Compared with the RT result, the gain decreased by about 40\%.
In order to specify the cause of this issue, we used the time-averaging function of the oscilloscope and compared the waveforms at the RT and with the simulation. The simulation methodology is discussed in the next section.

\begin{figure}[htbp]
  \begin{center}
   \includegraphics[width=110mm]{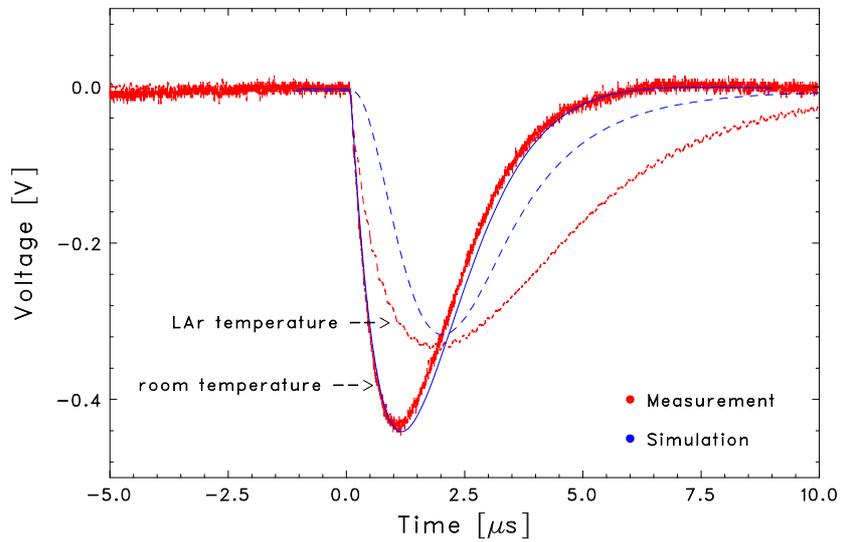}
  \end{center}
  \caption{Time-averaged waveforms at RT and LT. Data and simulation are shown in red and blue, respectively.}
  \label{fig:two_waveform}
 \end{figure}

\begin{figure}[htbp]
  \begin{center}
   \includegraphics[width=110mm]{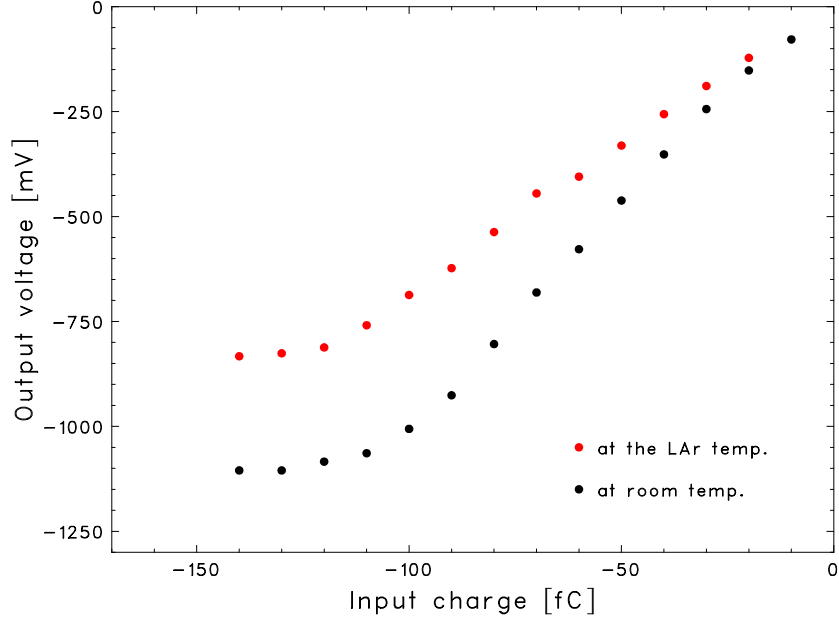}
  \end{center}
  \caption{Comparison of the dynamic range in the HG mode with $C_{\rm det}=300$~pF.}
  \label{fig:coldrange}
 \end{figure}

Figure \ref{fig:two_waveform} shows the time-averaged waveforms at RT and the LAr temperatures with an input charge of -40 fC and $C_{\rm det}=300$~pF. Compared with the RT results, not only the conversion gain, but also the peaking time was clearly affected by the temperature; there was an increase of ~60\% in the peaking time. The conversion gain at LT was determined to be 6.6 mV/fC, while the peaking time was about 1.6~$\mu$s. 
 This result was contrary to our expectations since the charge carrier mobility in silicon generally increases with decreasing temperature, while the thermal noise decreases at the same time in the LT environment. Figure \ref{fig:coldrange} shows the dynamic range at the LAr temperature.

\section{Possible cause of the Performance degradation and Simulation methodology for cold-electronics}
The analog performance at RT satisfied the requirements from the experiments, however, further optimization of the circuits and devices is clearly necessary for operation of circuits immersed in the LAr. In this section, we discuss possible causes of the performance degradation at LT and a unique simulation methodology for reliable cold electronics. 

\begin{figure}[htbp]
  \begin{center}
   \includegraphics[width=150mm]{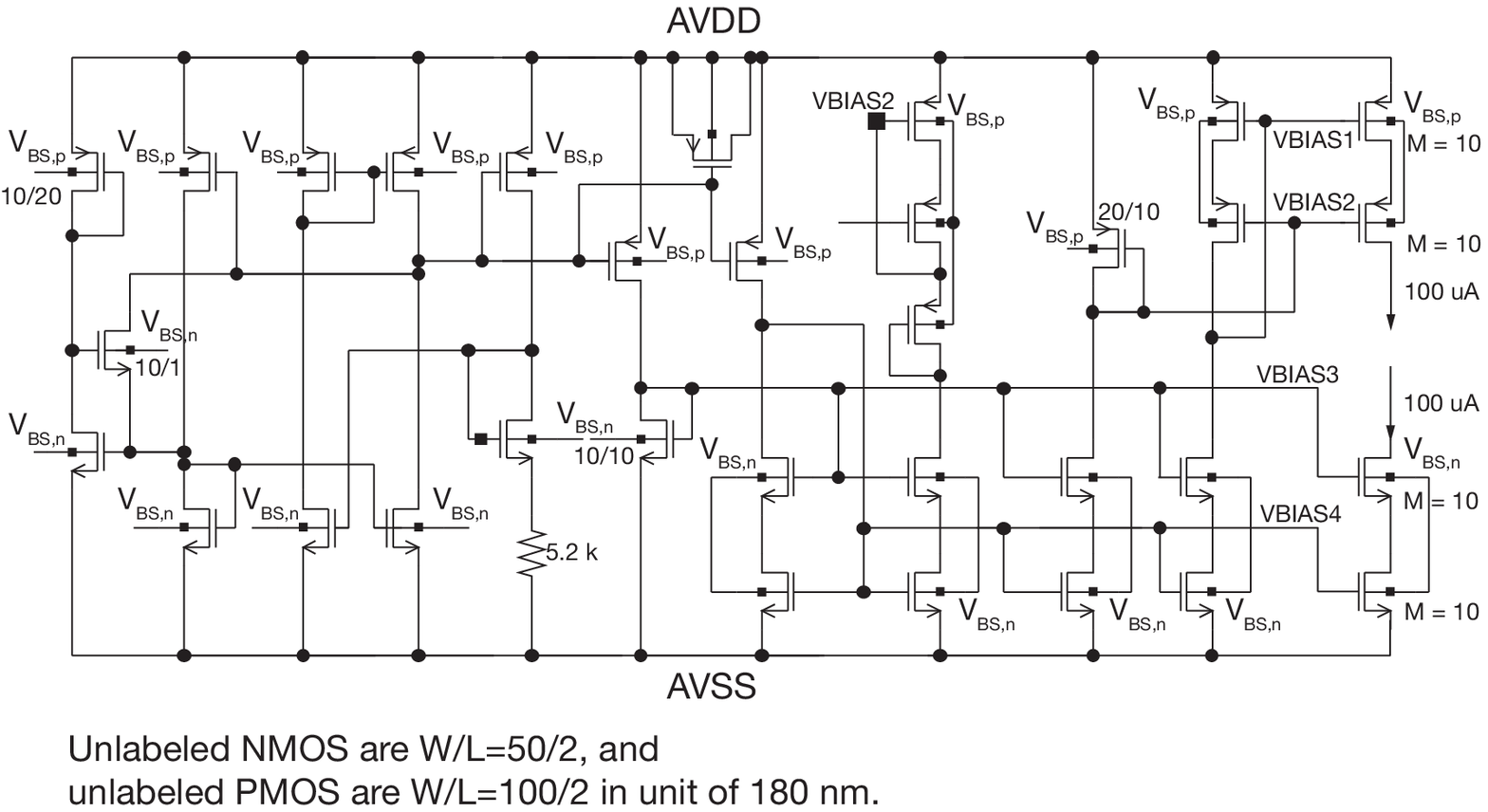}
  \end{center}
  \caption{Schematic of the improved bias circuit based on the beta-multiplier \cite{baker}. All substrate biases are separated from supply powers for the body-effect simulation.}
  \label{fig:beta_multiplier}
 \end{figure}

The gain and peaking time of the analog output are related to the operating points of the devices. It is generally known that the threshold voltage of the transistors increases as temperature is reduced, with a similar shifting-magnitude for n- and p-channel transistors, \eg, approximately 1~mV/K \cite{clark}. The SPICE parameters are supported at -40$^\circ$C from the vendor; however, there are hurdles to extending this model down to the LAr temperature. Instead of modifying the individual transistor parameters, we attempt to utilize the body effect by changing the substrate voltage. The threshold voltage, $V_{\rm th}$, of the transistor including the body effect is described with the Fermi potential of  bulk silicon $\Phi_{\rm F}$ with respect to the intrinsic Fermi level as 

\begin{equation}
V_{\rm th}=V_{\rm th0}+\gamma (\sqrt{|2\Phi_{\rm F}+V_{\rm SB}|} - \sqrt{|2\Phi_{\rm F}|}),
\end{equation}
where $V_{\rm th0}$ is the threshold voltage of the transistor at RT, $V_{\rm SB}$ is the source-bulk potential difference, and $\gamma$ is the body effect coefficient which typically lies in the range of 0.3 to 0.4~V$^{1/2}$ \cite{razavi}. Thus, it is possible to mimic the threshold voltages at LT, simply by changing $V_{\rm SB}$ without considering process-dependent parameters. In the actual simulation, we separated the substrate nodes from the source voltages, applying the negative values from the source voltage $V_{\rm BS, n}$ for n-channels and positive values of $V_{\rm BS,p}$ for p-channels (see Fig. \ref{fig:beta_multiplier} as a schematic example).

\begin{figure}[htbp]
 \begin{minipage}{0.5\hsize}
 \begin{center}
 \includegraphics[width=70mm]{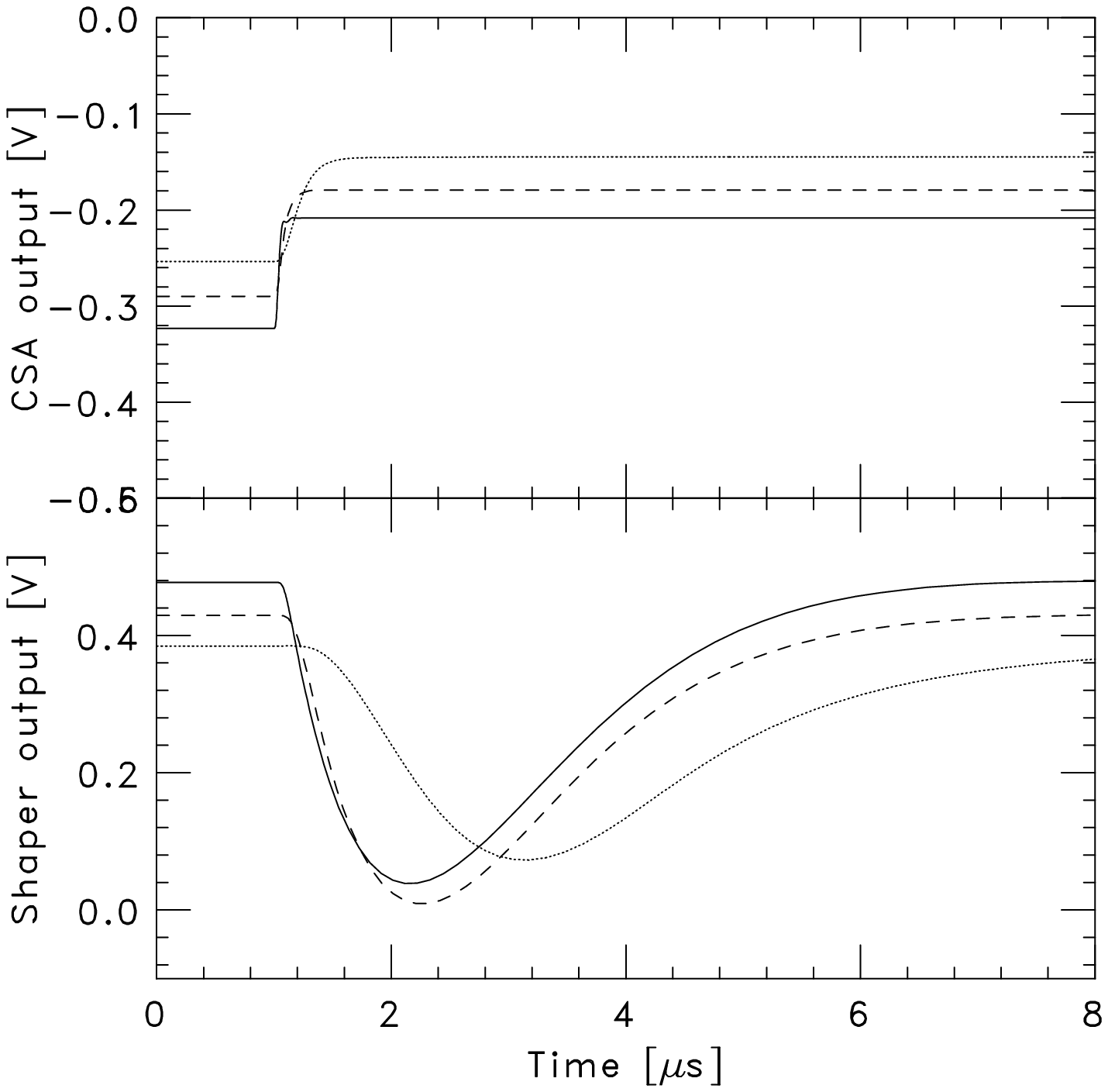}
 \end{center}
 \end{minipage}
 \begin{minipage}{0.5\hsize}
 \begin{center}
 \includegraphics[width=70mm]{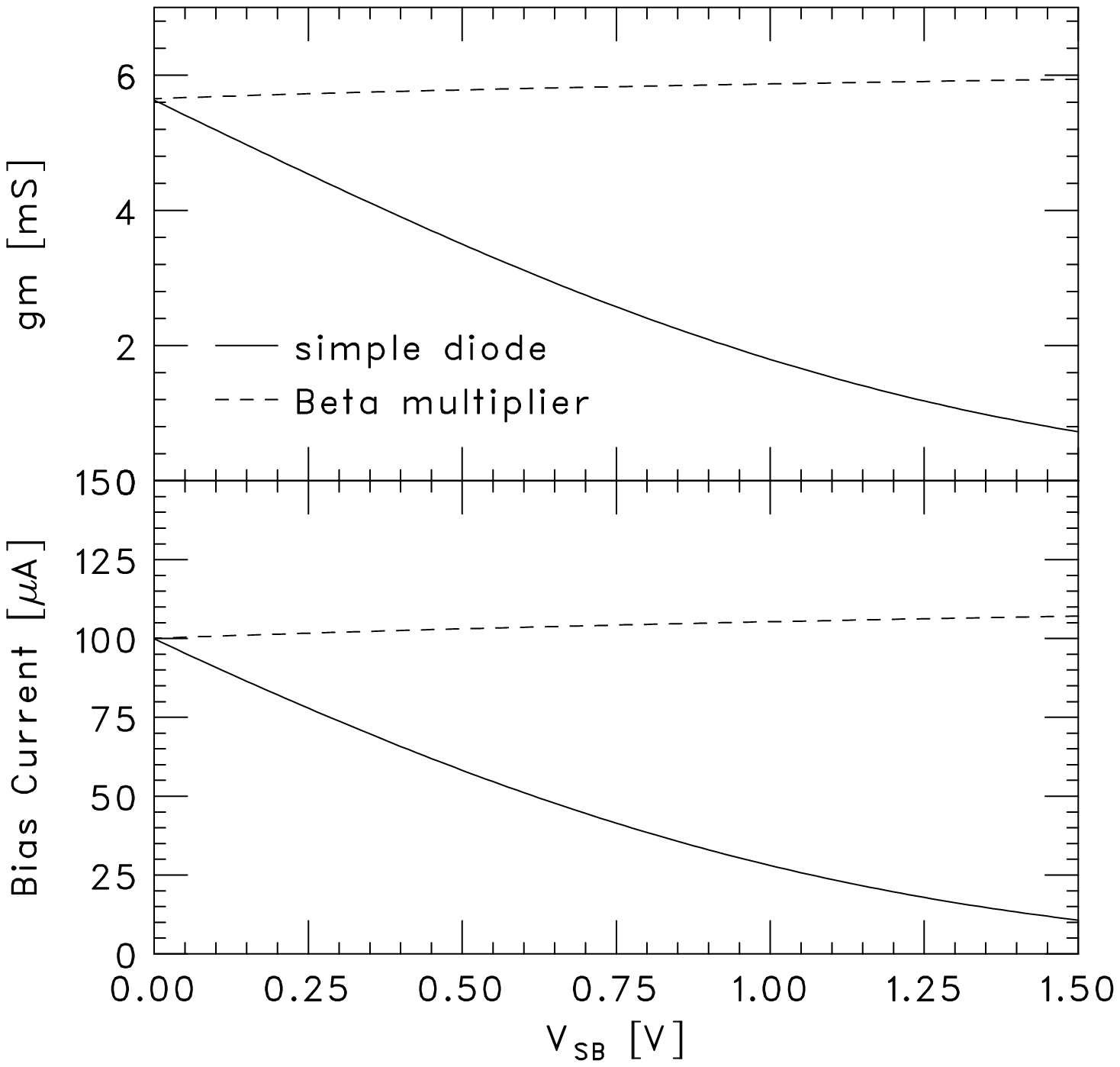}
 \end{center}
 \end{minipage}
 \caption{(Left): Simulation waveforms (top: CSA, bottom: shaper) with a typical process corner and different $V_{\rm SB}$ values. The injected charge is 40~fC in the HG mode. The baseline offset is also caused by the threshold shift. (Right): the transconductance of the input transistor (top) and internal reference bias current (bottom) as a function of $V_{\rm SB}$.}
 \label{fig:LTsim}
\end{figure}

The simulated waveforms with various $V_{\rm SB}$ are shown in the left of Figure \ref{fig:LTsim}. The top and bottom panels respectively show the outputs of the CSA and band-pass filter. As $V_{\rm SB}$ increases, the baselines at the CSA and band-pass filter outputs linearly shift from the RT condition, while the the rising edge of the CSA simultaneously becomes slower. On the other hand, the peaking time becomes longer and pulse height becomes lower. The threshold shift is estimated as ~200~mV if we assume the LAr temperatures, and thus, the corresponding $V_{\rm SB}$ is 0.88--1.35~V. Here we assumed  $2\Phi_{\rm F}=$0.8--0.9, and is temperature independent in this technology. We overlaid the simulation result of $V_{\rm SB}=1.2$~V in Fig. \ref{fig:two_waveform}. For direct comparison, we optimized offsets in the baseline and peaking-times. While the gain and peaking-time shifts are close to the measurements, the overall breadth of the measurement at LT is wider than that of the simulation. Although there is still room for bias tuning, this is plausibly caused by a unity-gain buffer implemented on the PCB board.

The cause of the waveform shift was found to be due to the linear decrease of the transconductance $g_m$ of the input FET. The right of Figure \ref{fig:LTsim} shows $g_m$ as a function of $V_{\rm SB}$. 
The rise time of the CSA ($t_{\rm r, CSA}$) is given as 
\begin{equation}
t_{\rm r, CSA}=\frac{C_{\rm det}}{g_{\rm m}} + \frac{C_{\rm L}}{\mu_0 \cdot g_{\rm m}},
\end{equation}
where $\displaystyle{\mu_0 = \frac{C_{\rm f}}{C_{\rm f}+C_{\rm det}}}$ 
and $C_{\rm L}$ is the load capacitance at the CSA output. We can see that the rise time of the CSA becomes larger as $g_{\rm m}$ decreases. The peaking time and pulse height of the filter can be explained as a consequence of the slow rising edge of the CSA. The decrease of $g_{\rm m}$  occurs due to the decrease of the internal bias current, which flows 100~$\mu$A at RT . The right side of Fig. \ref{fig:LTsim} shows the internal bias current as a function of $V_{\rm SB}$. 
As described in Section 2, we used a diode- and current mirror- configurations as a bias generator. However, even if the IBIAS node is connected to the supply power, the internal current deviates as temperature decreases due to the higher threshold voltage in the simple diode-configuration. As a result, the nominal 510~$\mu$A for the input FET is not provided properly, and consequently, the $g_m$ deteriorates as temperature decreases.

To reduce the threshold shift at the LAr temperature, we considered two approaches. The first is to apply a forward bias voltage, \eg, 0.5~V, to the source substrate junction. The resulting forward current might be negligible at low temperature because of the decrease in the intrinsic carrier concentration. This approach requires additional power supplies for substrate biasing, however, it is easier to tune the bias current externally. The second approach uses a feedback-based current circuit proposed in \cite{baker}.
Figure \ref{fig:beta_multiplier} shows a schematic of such bias circuit, based on the beta-multiplier. The addition of the resistor kills the closed loop gain, and the positive feedback system can be stable as long as its closed loop gain is less than one. The bottleneck of the circuit is that the gain of the loop increases as the size of resistor decreases. This pushes the feedback system closer to the instability. If the resistor, for example, is bonded out off-chip to set the current, it is likely that this bias circuit will oscillate, and thus, the circuit is basically self-biased with an on-chip resistor. For comparison, current output as a function of  $V_{\rm SB}$ is overlaid in the right of Fig. \ref{fig:LTsim} in a dashed line. We can see that the internal 100~$\mu$A is stably provided by this circuit.

\section{Conclusion}
We have newly developed the TPC readout chip in the 180~nm CMOS technology. The front-end ASIC is targeted at dual-phase LAr-TPCs for neutrino experiments and NI $\mu$-TPC for directional dark matter searches. We optimized the transistor parameters of the previous prototype chip, taking care in the layout for low-noise performance. The ENC reached  2695$\pm$71~e$^-$ (rms) for a 300~pF detector capacitance with a noise slope of 3.09 ~e$^-$/pF at RT. The dynamic range and shaping times also satisfy the experimental requirements. In the LAr temperature testing, we have acquired the analog waveforms from the ASIC, however, the noise level unexpectedly deteriorated with a longer shaping time. By comparing our results with the SPICE simulation, this issue was found to be caused by threshold voltage shifts of the transistors at LT. This work describes a unique simulation methodology for reliable cold electronics, which utilizes the body effect to mimic the threshold shifts. This method can be applicable to general cold electronics designs.

\acknowledgments
This work was supported by KAKENHI Grant-in-Aids (16H02189, 26104005, 17H01134, 18K03684, 19H05806), JSPS Bilateral Collaborations (Joint Research Projects and Seminars) program.


\begin{thebibliography}{99}
\bibitem{NEWAGE} T. Ikeda, et al., \emph{Results of a directional dark matter search from the NEWAGE experiment}, {\emph J. Phys.: Conf.} {\bf vol. 1468} (2020) 012042.

\bibitem{CYGNUS} K. Miuchi, et al., \emph{CYGNUS}, {\emph J. Phys.: Conf.} {\bf vol. 1468} (2020) 012044.

\bibitem{nuexp} A.~Badertscher, T.~Hasegawa, T.~Kobayashi, A.~Marchionni, A.~Meregaglia, T.~Maruyama, K.~Nishikawa and A.~Rubbia,
 \emph{A Possible Future Long Baseline Neutrino and Nucleon Decay Experiment with a 100 kton Liquid Argon TPC at Okinoshima using the J-PARC Neutrino Facility,}  arXiv:0804.2111 [hep-ph], 
DUNE collaboration, R.~Acciarri et al., Long-Baseline Neutrino Facility (LBNF) and Deep Underground Neutrino Experiment (DUNE), arXiv:1601.02984.

\bibitem{tanimori}T.~Tanimori et al., \emph{Detecting the WIMP-wind via spin-dependent interactions}, {\emph Phys. Lett. B} {\bf vol. 578} (2004) pg. 241--246.

\bibitem{mayet} F.~Mayet et al., \emph{A review of the discovery reach of directional Dark Matter detection}, {\emph Phys.~Rept} {\bf vol. 627} (2016) pg. 1--49.

\bibitem{minority} D.P.~Snowden-Ifft, {\emph Rev. Sci. Instrum.} {\bf vol. 85} (2014) 013303.

\bibitem{SF6_Ikeda2020} T. Ikeda, submitted for {\emph JINST} arXiv:2004.09706.


\bibitem{WA105} I. De Bonis, et al., Technical Design Report for large-scale neutrino detectors prototyping and phased performance assessment in view of a long-baseline oscillation experiment, Tech. Rep. CERN-SPSC- 2014-013. SPSC-TDR-004 (Apr 2014).
URL https://cds.cern.ch/record/1692375 .

\bibitem{LEM} A.~Badertscher, L.~Knecht, M.~Laffranchi, A.~Marchionni, G.~Natterer, P.~Otiougova et al., {\emph Construction and operation of a Double Phase LAr Large Electron Multiplier Time Projection Chamber}, in Proceedings of the Nuclear Science Symposium, Medical Imaging Conference and 16th International Workshop on Room-Temperature Semiconductor X-Ray and Gamma-Ray Detectors (NSS/MIC 2008/RTSD 2008), Dresden, Germany, 19--25 October 2008, pp. 1328--1334, arXiv:0811.3384.


\bibitem{nakazawa}  M.~Nakazawa et al., ``Prototype Analog Front-end for Negative-ion Gas and Dual-phase Liquid-Ar TPCs'', Jour. Instr., vol. 14, T01008, 2019.

\bibitem{miuchi} K.~Miuchi, \emph{Performance of the TPC with Micro Pixel Chamber Readout: micro-TPC}, {\emph IEEE~Trans.~Nucl.~Sci.} {\bf vol. 50} (2003) pg. 825--830.

\bibitem{giomataris} Y.~Giomataris, Ph. Rebourgeard, J.P. Robert, and G. Charpak, \emph{MICROMEGAS: a high-granularity position-sensitive gaseous detector for high particle-flux environments}, {\emph Nucl.~Instrum.~Methods~Phys.~Res.~A} {\bf vol. A376} (1996) pg. 29--35.



\bibitem{geronimo}  G. De Geronimo et al., {\emph``Front-end electronics for imaging detectors'', Nucl.~Instrum.~Methods~Phys.~Res.~A} {\bf vol. A471} (2001) pg. 192--199.

\bibitem{FPGA}https://www.xilinx.com/products/silicon-devices/fpga/artix-7.html.



\bibitem{SITCP}
T. Uchida, ``Hardware-Based TCP Processor for Gigabit Ethernet'',
IEEE TNS. {\bf 55}, 3 (2008).

\bibitem{clark} W. F. Clark et al., {\emph Low Temperature CMOS--A Brief Review}, IEEE trans. compon. hybrids mauf. technol., {\bf vol. 15} (1992) pg. 397--404.

\bibitem{razavi}B.~Razavi, {\emph Design of Analog CMOS Integrated Circuits}, McGrawHill.

\bibitem{baker}R.~J. Baker, {\emph CMOS Circuit Design, Layout, and Simulation}, Wiley.

\end{thebibliography}
\end{document}